\documentclass[pra, aps, superscriptaddress, twocolumn, showpacs, 10pt, floatfix]{revtex4-1}
\usepackage[utf8]{inputenc}
\usepackage[T1]{fontenc}
\usepackage[english]{babel}
\usepackage{amsmath}
\usepackage{amsfonts}
\usepackage{amssymb}
\usepackage{array}
\usepackage{xcolor}
\usepackage{graphics,graphicx}
\usepackage{epstopdf}
\newcommand{\e}{\mathrm{e}}
\newcommand{\opt}{\text{opt}}

\begin{document}
\title{Optimization of multiplexed single-photon sources\\ operated with photon-number-resolving detectors}
\author{Ferenc Bodog}
\affiliation{Institute of Physics, University of P\'ecs, Ifj\'us\'ag \'utja 6, H-7624 P\'ecs, Hungary}
\author{Matyas Mechler}
\affiliation{Institute of Physics, University of P\'ecs, Ifj\'us\'ag \'utja 6, H-7624 P\'ecs, Hungary}
\author{Matyas Koniorczyk}
\affiliation{Institute for Solid State Physics and Optics, Wigner Research Centre for Physics,\\ P.O. Box 49, H-1525 Budapest, Hungary}
\author{Peter Adam}
\email{adam.peter@wigner.mta.hu}
\affiliation{Institute of Physics, University of P\'ecs, Ifj\'us\'ag \'utja 6, H-7624 P\'ecs, Hungary}
\affiliation{Institute for Solid State Physics and Optics, Wigner Research Centre for Physics,\\ P.O. Box 49, H-1525 Budapest, Hungary}

\date{\today}

\begin{abstract}
Detectors inherently capable of resolving photon numbers have undergone a significant development recently, and this is expected to affect multiplexed periodic single-photon sources where such detectors can find their applications. We analyze various spatially and time-multiplexed periodic single-photon source arrangements with photon-number-resolving detectors, partly to identify the cases when they outperform those with threshold detectors. We develop a full statistical description of these arrangements in order to optimize such systems with respect to maximal single-photon probability, taking into account all relevant loss mechanisms. The model is suitable for the description of all spatial and time multiplexing schemes. 
Our detailed analysis of symmetric spatial multiplexing identifies a particular range of loss parameters in which the use of the new type of detectors leads to an improvement. Photon number resolution opens an additional possibility for optimizing the system in that the heralding strategy can be defined in terms of actual detected photon numbers. Our results show that this kind of optimization opens an additional parameter range of improved efficiency. Moreover, this higher efficiency can be achieved by using less multiplexed units, i.e., smaller system size as compared to threshold-detector schemes.
We also extend our investigation to certain time-multiplexed schemes of actual experimental relevance. We find that the highest single-photon probability is 0.907 that can be achieved by binary bulk time multiplexers using photon-number-resolving detectors.
\end{abstract}

\maketitle

\section{Introduction}\label{Sec:intro}

Construction of periodic single-photon sources (PSPS) is subject to intensive research due to the numerous applications of such devices in quantum information processing~\cite{Knill2001, Kok2007, Gisin2002, Scarani2009, Duan2001, Sangouard2011, Bennett1993, Bouwmeester1997, Merali2011, Koniorczyk2011, Spring2013, Broome2013, Tillmann2013} and photonic quantum technology~\cite{Gerry1999, Lund2004, He2009, Adam2010, Lee2012}. The most promising realization of PSPS are the heralded single-photon sources (HSPS) based on heralding one member (termed as the signal) of a correlated photon pair generated in nonlinear optical media by detecting the other member (referred to as the idler) of the photon pair.

The most widely used processes applied for pair generation are spontaneous four-wave mixing (SFWM) in optical fibers \cite{Smith2009,Cohen2009,Soller2011,MScott2015} and
spontaneous parametric down-conversion (SPDC) in bulk crystals \cite{Mosley2008, Zhong2009, Evans2010, Brida2011, Broome2011} or waveguides \cite{Fiorentino2007, Eckstein2011, Horn2013}. The latter process can yield highly indistinguishable single photons in an almost ideal single mode with known polarization \cite{Mosley2008,Evans2010, Eckstein2011,Fortsch2013}.
However, the probabilistic nature of photon pair generation in nonlinear sources results in a limit of the achievable single-photon probability. For example, for weaker spectral filtering, the theoretically achievable maximal single-photon probability of HSPS is 0.367 which corresponds to the single-photon probability of Poissonian statistics. For  a sufficiently narrow filtering ensuring the photon indistinguishability, the photon statistics of the pairs is thermal resulting in a lower single-photon probability of 0.25.
To overcome this issue and for enhancing the single-photon probability while simultaneously suppressing multiphoton noise in HSPS, various techniques of multiplexing, namely spatial multiplexing \cite{Migdall2002, Shapiro2007, Ma2010, Collins2013, Meany2014, Francis2016, KiyoharaOE2016} and time multiplexing \cite{Pittman2002, Jeffrey2004, Mower2011, Schmiegelow2014, Francis2015, Kaneda2015, Rohde2015, XiongNC2016, Hoggarth2017, HeuckNJP2018, Kaneda2019, Lee2019, MagnoniQIP2019} were proposed in the literature.

The principle of multiplexing is to reroute heralded photons generated in a set of multiplexed units realized in time or in space to a single output mode by a switching network.
The mean photon number of the generated photon pairs in a multiplexed unit should be kept low ensuring that the multi-photon events are negligible while the overall probability of successful heralding in all the multiplexed units should be high ensuring a high single-photon probability.

In the case of spatial multiplexing several individual pulsed HSPS are used in parallel. 
After a successful heralding event in the idler arm of a source (i.e. in a multiplexed unit), the corresponding signal photon is directed by binary photon routers (2-to-1) to a single output.
The system of routers can be either symmetric (log-tree) or asymmetric (chain)  \cite{MazzarellaPRA2013,BonneauNJP2015}.
The periodicity of such PSPS stem from the period of the pulsed pump laser. Spatial multiplexing has been successfully demonstrated experimentally up to four multiplexed units by using SPDC in bulk crystals \cite{Ma2010,KiyoharaOE2016} and waveguides \cite{Meany2014}, and up to two multiplexed units by using SFWM in photonic crystal fibers \cite{Collins2013, Francis2016}.

The idea of time multiplexing is based on the detection of idler photons of a pulsed or continuous nonlinear photon pair source in time slots within a time period. 
Hence, a multiplexed unit is a possible time slot in this case. 
In case of a successful detection event the heralded signal photons are delayed to leave the system at the end of the given time period resulting in a periodic single-photon source.
Timing of the arrival of the output photons can be controlled by a switchable optical storage cavity or loop \cite{Pittman2002, Jeffrey2004, Rohde2015, HeuckNJP2018}, or using a binary division strategy \cite{Mower2011,Schmiegelow2014,MagnoniQIP2019, Lee2019}.
Time-multiplexing based on a delay loop has been realized experimentally in a fiber-integrated system using SFWM up to four time slot \cite{XiongNC2016,Hoggarth2017}.
Optical storage cavities and SPDC sources were used in experiments realizing PSPS via large-scale time multiplexing up to 40 time slots \cite{Kaneda2015, Kaneda2019}.
The highest single-photon probability that has been achieved in these latter experiments is 0.667, which is the highest one realized in multiplexed PSPS until now \cite{Kaneda2019}.
Moreover, the generated photons collected into a single-mode fiber were highly indistinguishable.

In principle, in an ideal lossless multiplexed system the single-photon probability asymptotically tends to one by increasing the number of multiplexed units and simultaneously decreasing the mean photon number of the incoming photon pairs.
However, losses of non-ideal optical elements in the heralding stage as well as in the multiplexing system impose a limitation on the performance of multiplexed PSPS~\cite{MazzarellaPRA2013, BonneauNJP2015}.
Therefore a full statistical treatment taking into account all relevant loss mechanisms is required for the analysis of such systems.
In Ref.~\cite{Adam2014} such a theoretical framework was presented for the description and optimization of spatially and time-multiplexed systems built up with threshold (or on-off) detectors.

The optimization revealed that for a given set of losses there is a number of multiplexed units and a mean photon number of photon pairs in each multiplexed unit for which the single-photon probability is maximal.
The analysis also showed that the highest single-photon probability can be achieved by using bulk time multiplexers based on binary division strategy.
Using experimentally realizable optical elements in this system the achievable single-photon probability is $.85$.
In Ref.~\cite{Bodog2016} the theoretical framework was extended to describe combined multiplexers that apply both time and spatial multiplexing in a single setup \cite{GlebovAPL2013, LatypovJPCS2015, MendozaOptica2016}.
The optimization showed that the combination of the two types of multiplexing can lead to a decrease in the number of required nonlinear sources and a possible increase in the achievable repetition rate of the system as compared to the standalone use of the optimized spatial or time multiplexer, while maintaining the achieved single-photon probability.

Beside multiplexing, another way of avoiding the occurrence of multi-photon events at the heralding stage of these systems is the application of photon-number-resolving detection and considering the events of detecting exactly one photon.
Accordingly, single-photon detectors with number resolving capability (SPDs) were generally used in recent multiplexed PSPS experiments \cite{Collins2013, Kaneda2015, Francis2016, KiyoharaOE2016, XiongNC2016, Hoggarth2017, Kaneda2019}.
A possible realization of SPDs is based on simultaneous use of threshold detectors with either a temporal or spatial multiplexing scheme \cite{PaulH1996, PaulH1997,KrusePRA2017, ChenSR2017, JoenssonPRA2019}, which leads to a limited detection efficiency.

Meanwhile, developing high-efficiency photon-number-resolving detectors (PNRDs) is of great research interest due to their various applications in photonic quantum technologies.
The best known realizations of such devices are transition edge sensors \cite{Lita2008, Lita2009, Fukuda2011}, quantum dot optically gated field-effect transistors \cite{Gansen2007, Kardynal2007}, superconducting nanowire detectors~\cite{Divochiy2008, Jahanmirinejad2012}, and fast-gated avalanche photodiodes~\cite{Kardynal2008, Thomas2010}.
Detector efficiencies as high as 98\%~\cite{Fukuda2011} have already been reported with almost ideal photon-number discrimination at low photon numbers using transition edge sensors in the near-infrared regime.

The progress in the experimental realization of photon detection with photon-number-resolving capability naturally inspires the development of a theoretical framework for multiplexed PSPS  equipped with PNRDs. As we saw before, in case of multiplexed PSPS employing threshold detectors, a proper theoretical framework leads to a detailed understanding as well as valuable information for the design of the system. Motivated by these, in the present paper we develop the full statistical description of such systems including all relevant loss mechanisms. 
We incorporate into the model the use of PNRDs with arbitrary detection strategy, that is, detecting an arbitrary set of predefined number of photons in the idler arm for which the generated signal photons are allowed to enter the multiplexer.
This model allows for the maximization of single-photon probabilities of multiplexed PSPS under realistic experimental conditions via optimization in the number of multiplexed units and the mean photon number of photon pairs in each of them.
We compare the achievable single-photon probabilities of spatially multiplexed systems equipped with PNRDs with the ones containing threshold detectors over a wide range of the parameters describing the various losses. This will make it possible to predict when the use of PNRDs really pays off.
We accomplish this comparison for relevant time multiplexed systems for selected sets of experimentally feasible parameters. We do this with the aim of identifying the feasible arrangement which is likely to prove the most efficient.

The paper is organized as follows: in Sec.~\ref{Sec:overview} we give a short overview of spatial and time multiplexing. In Sec.~\ref{Sec:math} we introduce the theoretical framework that can be used to carry out the optimization and analysis of multiplexed PSPS. In Sec.~\ref{Sec:result} we present the results of the optimization for relevant spatially and time-multiplexed systems. Finally, we conclude in Sec.~\ref{Sec:conc}.

\section{Spatial and Time Multiplexing}\label{Sec:overview}

In this section we briefly review spatial and time multiplexed PSPS. The focus will be on those schemes which bear experimental relevance and our model is applied to in the rest of this paper.

\begin{figure}[!bt]
\includegraphics[width=\columnwidth]{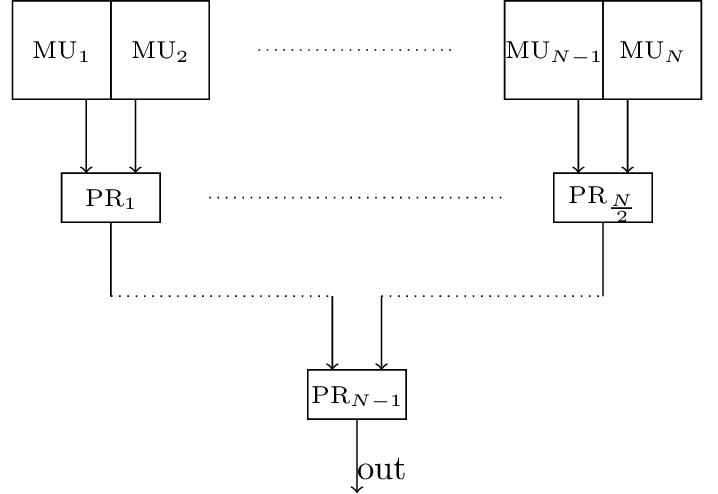}
\caption{\label{Figure1} Schematics of spatially multiplexed PSPS with a symmetric (log-tree) spatial multiplexer. MU$_i$ denotes the $i$th multiplexed unit, the PR$_j$s are binary (2-to-1) photon routers.}
\end{figure}

Figure~\ref{Figure1} shows a schematic depiction of a spatially multiplexed PSPS with a symmetric (log-tree) spatial multiplexer most frequently used in experiments \cite{Ma2010,Meany2014}.
In the figure MU$_i$ denotes a multiplexed unit where a signal photon can enter the multiplexer given that its twin partner, that is, the idler photon of a photon pair coming from a pulsed nonlinear source is detected.
In this case the spatial multiplexer built up with binary (2-to-1) photon routers (PR) in a log-tree arrangement where the $N$ routers are arranged in $m=\log_2 N$ levels.
After a successful heralding event in one of the multiplexed units, the switching network is actively controlled to route the heralded photon to the common output.

In this scheme it is generally assumed that photon routers are symmetric, that is, the router transmission $V_r$ does not depend on the input ports of the individual routers \cite{Ma2010, Francis2015, BonneauNJP2015, MazzarellaPRA2013}. The total probability of transmission $V_n^{\rm SSM}$ for the $n$th multiplexed unit for such symmetric spatial multiplexers (SSM) with symmetric routers  reads  
\begin{equation}
V_n^{\rm SSM}=V_bV_r^{\log_2N},\label{eq:SSMLoss}
\end{equation}
where $N$ is the number of multiplexed units. $V_b$ is a general transmission coefficient independent of the number of multiplexed units, such as the collection efficiency at the heralding stage or the transmission of the possible pre-delay.

\begin{figure}[tb]
\includegraphics[width=\columnwidth]{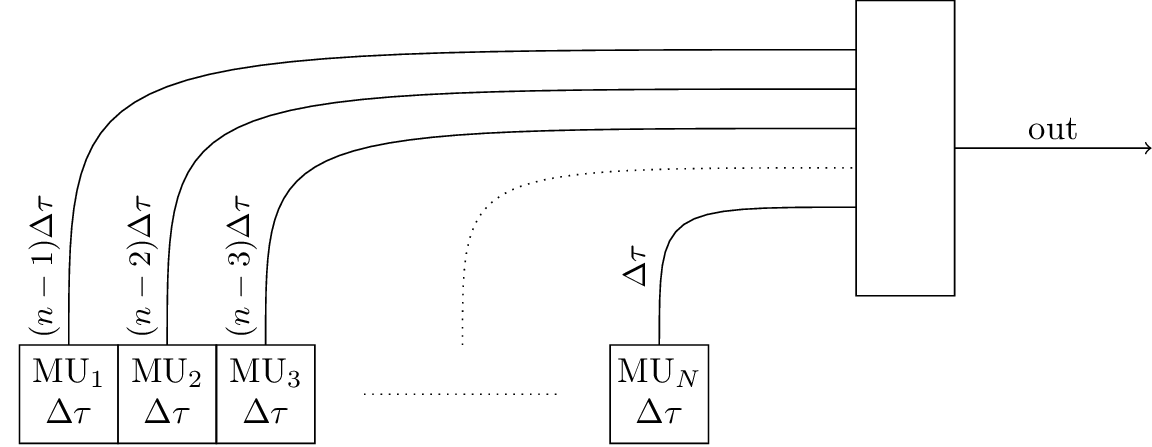}
\caption{\label{Figure2} Schematics of time multiplexed PSPS. 
MU$_i$ denotes the $i$th multiplexed unit, that is, a time slot in this case.
The heralded signal photon emerging from one of the multiplexed units is delayed to leave the system at the end of the expected time period $T=N\Delta\tau$.}
\end{figure}

Now let us turn our attention to time multiplexing. 
Figure~\ref{Figure2} demonstrates its principle.
The expected time period $T$ of the periodic single-photon source is divided into $N$ time slots of length $\Delta \tau$, which play the role of multiplexed units.
In these multiplexed units the idler part of the photon pairs emitted by a pulsed or continuous nonlinear source are detected.
After a successful heralding event in the $n$th time slot the heralded photon is delayed by the time $(N-n)\Delta \tau$ before it reaches the output of the PSPS.

One way to realize a controllable time delay is to use a switchable storage loop or storage cavity with a passage time $\Delta \tau$ \cite{Pittman2002, Jeffrey2004, Francis2015, Kaneda2015, Rohde2015, XiongNC2016, Hoggarth2017, HeuckNJP2018, Kaneda2019}.
The heralded photons generated in the $n$th time slot will thus pass through the loop/cavity $N-n$ times before they are released.
Denoting the transmission coefficient of a single cycle of the storage loop by $V_c$, the total transmission $V_n^{\rm LTM}$ corresponding to the $n$th time slot can be written as
\begin{equation}
V_n^{\rm LTM}=V_bV_c^{N-n}.\label{eq:LTMLoss}
\end{equation}
Here the general transmission coefficient $V_b$ characterizes the losses in the system independent of the number of multiplexed time slots, such as, e.g., the collection efficiency or the loss during the switching into and out from the storage loop.
We note that Eq.~\eqref{eq:LTMLoss} also describes the total transmission probability of the asymmetric spatial multiplexer proposed in Ref.~\cite{MazzarellaPRA2013}: the role of the single-cycle transmission $V_c$ is played by the router transmission in that case.

Another way of implementing a controllable time delay is to use a binary division strategy \cite{Mower2011,Schmiegelow2014,MagnoniQIP2019, Lee2019}.
In this case, the heralded photons are directed to switchable delay lines with different time lengths ($\Delta \tau$, $2\Delta \tau$, $4\Delta \tau\dots$ ) which are multiples of powers of 2. 
Signal photons heralded in the $n$th time slot travel only through those delay lines whose length corresponds to the nonzero digits in the binary representation of the time delay $(N-n)\Delta \tau$.
A possible realization of such a system in bulk optics was proposed in Ref.~\cite{Adam2014}.
The delay lines of this scheme contain Pockels cells and polarizing beam splitters with reflection and transmission efficiency $V_{\rm re}$ and $V_t$, respectively (see Figs. 3 and 4 in Ref.~\cite{Adam2014}). The total transmission probability $V_n^{\rm BTM}$ corresponding to the $n$th time slot for such a binary bulk time multiplexer reads
\begin{equation}
    V_n^{\rm BTM}=V_b V_{\rm re}^h V_t^{(l-h)} V_p^{(N-n)/N},\label{eq:BTMLoss}
\end{equation}
where $h$ is the Hamming weight of $N-n$ (the number of ones in its binary representation), and $l=\log_2N$.
The transmission coefficient $V_b$ merges the effects independent of the $n$th time slot.
The transmission coefficient $V_p$ characterizes the loss due to the propagation through the whole medium of the multiplexer, that is, the medium of the longest delay of the binary time multiplexer.

\section{Mathematical framework}\label{Sec:math}
In this section we present the statistical framework suitable for describing all kinds of multiplexed PSPS equipped with PNRDs. Throughout our calculations we consider a PNRD that is capable of distinguishing the number of detected photons up to a given value $J_b$. 

Consider a spatially or time multiplexed PSPS having $N$ multiplexed units, respectively. In the case of symmetric spatial multiplexing $N$ is a power of 2.
Assume that in the $n$th multiplexed unit $l$ photon pairs are generated by a nonlinear source and the input ports of the multiplexer are opened in case of detecting a predefined number of photons $j$ ($j\leq l$) during a heralding event. The probability that $i$ signal photons reach the output of the multiplexer in general reads
\begin{widetext}
\begin{eqnarray}
P_i^{(S)}=\big(1-\sum_{j\in S}P^{(D)}(j)\big)^N\delta_{i,0}+\sum_{n=1}^N\left[\big(1-\sum_{j\in S}P^{(D)}(j)\big)^{n-1}\times\sum_{l=i}^\infty\sum_{j\in S} P^{(D)}(j|l)P^{(\lambda)}(l)V_n(i|l)\right],\label{general_formula}
\end{eqnarray}
\end{widetext}
where $P^{(D)}(j)$ is the probability of detecting exactly $j$ photons arm in a multiplexed unit, $P^{(D)}(j|l)$ is the conditional probability of detecting $j$ photons given that $l$ photons reach the detector, $P^{(\lambda)}(l)$ is the probability of generating $l$ photon pairs, and $V_n(i|l)$ is the conditional probability that $i$ photons reach the output of the multiplexer given that $l$ signal photons arrive from the $n$th multiplexed unit into the system.
The set $S$ contains the predefined number of detected photons in a multiplexed unit for which the generated signal photons are allowed to enter the multiplexer. 
Hence, it describes the application of an optional heralding strategy that can be realized only by PNRDs. The set $S$ can be any subset of the set of positive integers $\mathbb{Z}^+$ up to $J_b$.
We note that for $S=\mathbb{Z}^+$ we obtain the same formulas we had introduced in Ref.~\cite{Adam2014} for describing multiplexed systems with threshold detectors, as PNRDs can be also used as threshold detectors by ignoring the number of detected photons.

The first term in Eq.~\eqref{general_formula} describes the case when none of the detectors have detected a photon number in $S$. This term contributes only to the probability $P_0^{(S)}$, that is, to the case where no photon reaches the output.
The second term describes the case when, even though there are $l$ photons entering the multiplexer from the $n$th multiplexed unit after heralding, only $i$ of these reach the output due to the losses of the multiplexer.

In Eq.~\eqref{general_formula} the conditional probability $P^{(D)}(j|l)$ that a detector with efficiency $V_D$ detects $j$ out of $l$ photons ($j \leq l$) in a multiplexed unit reads
\begin{eqnarray}
P^{(D)}(j|l)=\binom{l}{j}V_D^j(1-V_D)^{l-j}.
\end{eqnarray}
The total probability $P^{(D)}(j)$ of detecting $j$ photons out of the incoming ones can be written as
\begin{eqnarray}
P^{(D)}(j)=\sum_{l=j}^\infty P^{(D)}(j|l)P^{(\lambda)}(l)
\end{eqnarray}
where $P^{(\lambda)}(l)$ is the probability that $l$ photon pairs were generated by a nonlinear source. In our calculations  we assume that the probability distribution of pair generation $P^{(\lambda)}(l)$ is Poissonian, that is,
\begin{equation}
P^{(\lambda)}(l)=\frac{\lambda^l e^{-\lambda}}{l!},
\end{equation}
where $\lambda$ is the mean photon number of the generated photon pairs. 
We note that the validity of the calculation does not depend on the particular distribution. 
Hence, a thermal distribution can also be assumed here.

The conditional probability  $V_n(i|l)$ that $i$ signal photons reach the output of the multiplexer given that $l$ signal photons enter the multiplexer at the $n$th multiplexed unit can be calculated as
\begin{equation}
V_n(i|l)=\binom{l}{i}V_n^i(1-V_n)^{l-i}.
\end{equation}
The subscript $n$ is present for the possibility that the loss can depend on the actual multiplexed unit the photons have arrived at.
When analyzing a particular setup, the corresponding $V_n$ has to be substituted here, for instance, the one in  Eq.~\eqref{eq:SSMLoss}, Eq.~\eqref{eq:LTMLoss}, or Eq.~\eqref{eq:BTMLoss} for the schemes studied in this paper. 

Finally, we note that for threshold detectors ($S=\mathbb{Z}^+$) and SPDs ($S=\{1\}$), simple formulas can be derived from Eq.~\eqref{general_formula}. For instance, assuming a Poissonian distribution, for the single-photon probabilities $P_1^{(S)}$ we obtain
\begin{eqnarray}
P_{1}^{(\mathbb{Z}^+)}=&&\sum_{n=1}^\mathrm{N}\e^{-\lambda V_D(n-1)}\lambda V_n\e^{-\lambda}\nonumber\\&&\times(\e^{\lambda(1-V_n)}-(1-V_D)\e^{\lambda(1-V_n)(1-V_D)}),\label{eq:SPDformula}\\
P_{1}^{(1)}=&&\sum_{n=1}^\text{N}(1-V_D\lambda\e^{-V_D\lambda})^{n-1}\left[1+(1-V_D)(1-V_n)\lambda\right]\nonumber\\&&\times\lambda V_DV_n\e^{\left[(1-V_D)(1-V_n)-1\right]\lambda}.\label{eq:Thresholdformula}
\end{eqnarray}
Using the formula in Eq.~\eqref{general_formula}, the performance of an arbitrary multiplexed periodic single-photon source with arbitrary detector or arbitrary detection strategy with PNRDs can be analyzed in detail, including the optimization of the system to maximize the single-photon probabilities. 

\section{Optimization of multiplexed PSPS}\label{Sec:result}

In this section we present our results on the optimization of various multiplexed PSPS equipped with PNRDs. 
The goal is to achieve a maximal single-photon probability. 
Naively one would expect that the use of more advanced detectors will result in an overall performance improvement to the system, as compared to using the simpler (i.e. threshold) detectors. 
Having a description of both cases at hand, we can compare the performance of this system to that of the ones operated with threshold detectors. 
This analysis will result in nontrivial consequences.

\subsection{Symmetric spatial multiplexing with single-photon or threshold detectors}\label{sSec:spmsptd}
\label{sec:th}

\begin{figure}[tb]
\includegraphics[width=\columnwidth]{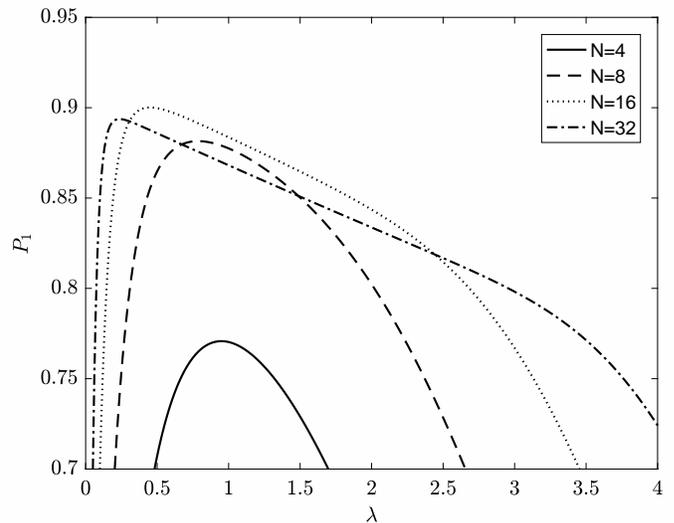}
\caption{\label{Figure3} The single-photon probability $P_1$ plotted against the input mean photon number $\lambda$ of a multiplexed unit for detector efficiency $V_D=0.95$ and router transmission $V_r=0.98$ for various values of the number of multiplexed units $N$. The result of the optimization: maximal single-photon probability $P_{1,\max}=0.9$, the optimal input mean photon number $\lambda_\opt=0.45$ and the optimal number of multiplexed unit $N_\opt=16$.}
\end{figure}

Let us first consider a PSPS based on symmetric spatial multiplexing and containing either single-photon or threshold detectors with detector efficiency $V_D$ in the idler arms of the nonlinear photon pair sources.
(Recall that by single-photon detector we mean a detector capable of identifying \emph{exactly one photon}, e.g. it can be a PNRD so that only the outcome corresponding to 1 photon is considered as a detection event.)
Using the framework presented in the previous section the optimization of the system consists in the following.
We fix the total transmission $V_n^{\text{SSM}}$ of the system, that is, the transmission coefficients $V_r$ and $V_b$, and the detector efficiency $V_D$. 
Recall that the spatial multiplexer under consideration
is built up from symmetric photon routers. Two parameters are left which can be considered as decision variables of the optimization procedure: the mean photon number $\lambda$ of the photon pairs generated by the nonlinear sources and arriving at the detectors in the multiplexed units, and the number $N$ of these units.
First, we calculate the single-photon probabilities $P_1$ as a function of the input mean photon number $\lambda$ for each reasonable value of the number $N$ of multiplexed units. 
Some instances of these functions are plotted in Fig.~\ref{Figure3}. 
Next, we determine the maximum values for each of these functions and select the optimal number $N_{\opt}$ of multiplexed units for which the maximal single-photon probability $P_1$ is the highest. 
The input mean photon number corresponding to this maximal single-photon probability $P_{1,\max}$ is the optimal mean photon number $\lambda_{\opt}$.

\begin{figure}[tb]
\includegraphics[width=\columnwidth]{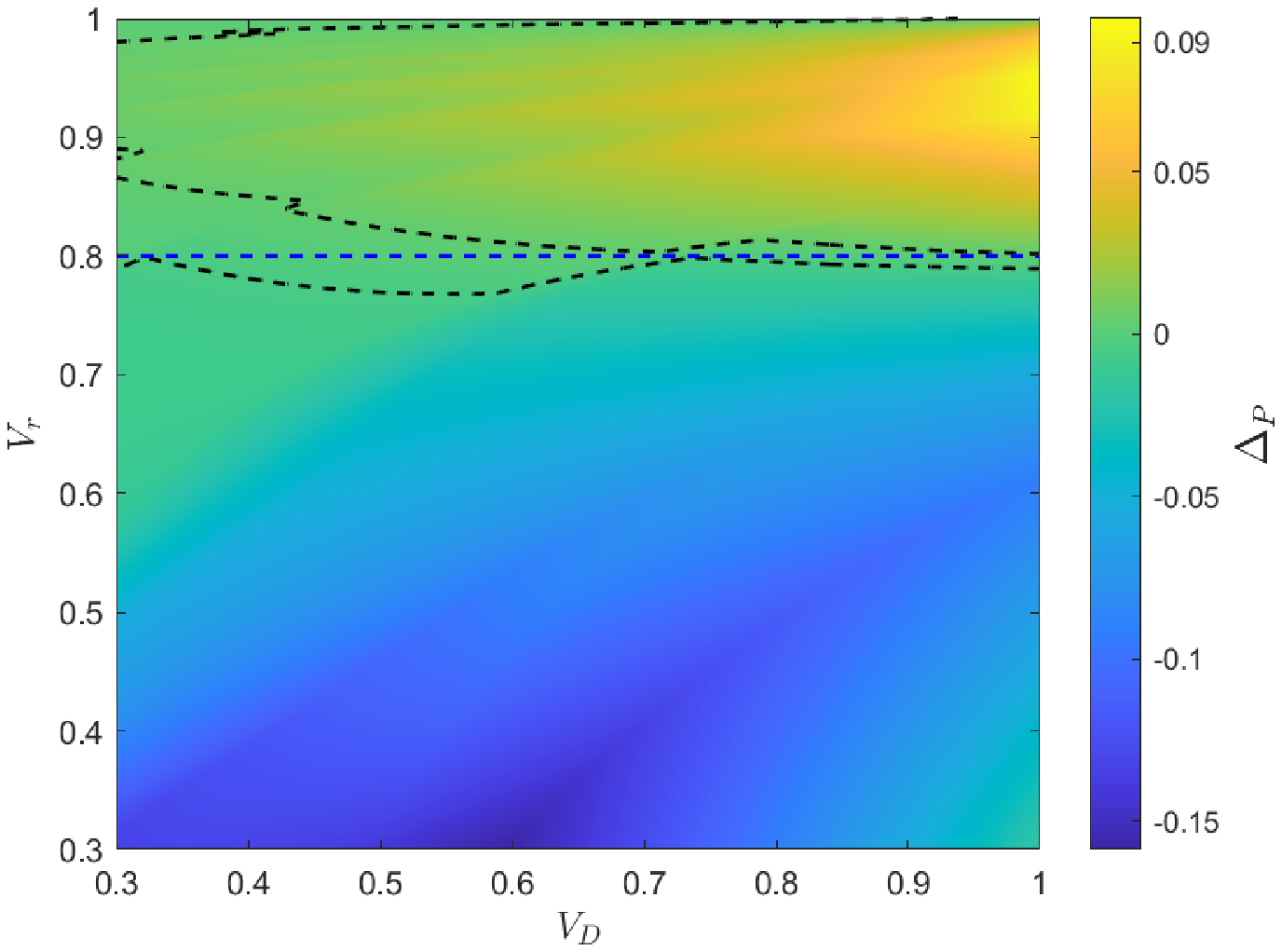}
\caption{\label{Figure4} The difference $\Delta_P=P_{1,\max}^{\text{SPD}}-P_{1,\max}^{\text{Th}}$ between the maximal single-photon probabilities for PSPS based on symmetric spatial multiplexing with two types of detectors: for those with SPDs and for those with threshold detectors, as a function of the router transmission $V_r$ and detector efficiency $V_D$. In the regions bounded by dashed lines around $V_r\approx 1$ and $V_r \approx 0.8$ the absolute difference $|\Delta_P|< 0.004$.}.
\end{figure}

\begin{small}
\begin{table*}[tb]
\caption{Optimal PSPS based on symmetric spatial multiplexing with SPDs. Maximal single-photon probabilities $P_{1,\max}$ and the required number of multiplexed units $N_{\opt}$ and input mean photon numbers $\lambda_{\opt}$ at which they can be achieved, calculated for different $V_r$ router transmissions and five different values of the detector loss $V_D$.\label{tab:PNRD1}}
\begin{ruledtabular}
\begin{tabular}{c| c@{\;}cc|c@{\;}cc|*2{c@{\;}cc|}c@{\;}cc}
      &\multicolumn{3}{c|}{$V_D=0.3$}&\multicolumn{3}{c|}{$V_D=0.6$}&\multicolumn{3}{c|}{$V_D=0.8$}&\multicolumn{3}{c|}{$V_D=0.9$}&\multicolumn{3}{c}{$V_D=0.98$}\\\hline
$V_r$ &
$N_{\text{opt}}$&$P_{1,\max}$&$\lambda_{\text{opt}}$&
$N_{\text{opt}}$&$P_{1,\max}$&$\lambda_{\text{opt}}$&
$N_{\text{opt}}$&$P_{1,\max}$&$\lambda_{\text{opt}}$&
$N_{\text{opt}}$&$P_{1,\max}$&$\lambda_{\text{opt}}$&
$N_{\text{opt}}$&$P_{1,\max}$&$\lambda_{\text{opt}}$\\\hline
0.30&
2&	 0.236&	 3.117&
1&	 0.221&	 1.000&
1&	 0.294&	 1.000&
1&	 0.331&	 1.000&
1&  0.361&	 1.000 \\
0.40&
4	 &0.290	 &4.754 &		
2	 &0.258	 &1.704 &		
1	 &0.294	 &1.000 &		
1	 &0.331	 &1.000 &		
1&  0.361&	 1.000		\\ 
0.50&	 
8	 &0.326	 &6.482 &       
2	 &0.288	 &1.515 &       
2	 &0.298	 &1.228 &       
1	 &0.331	 &1.000 &       
1&  0.361&	 1.000 \\
0.55&	 
8	 &0.352	 &5.510 &       
4	 &0.309	 &2.088 &       
2	 &0.320	 &1.195 &       
1	 &0.331	 &1.000 &       
1&  0.361&	 1.000 \\
0.60&	 
8	 &0.368	 &4.384 &       
4	 &0.341	 &1.870 &       
2	 &0.342	 &1.162 &       
2	 &0.352	 &1.078 &       
1&  0.361&	 1.000 \\
0.65&	 
8	 &0.380	 &3.256 &       
4	 &0.371	 &1.655 &       
4	 &0.366	 &1.309 &       
2	 &0.377	 &1.063 &       
2&  0.388&	 1.012 \\
0.70&	 
8	 &0.393	 &2.337 &       
4	 &0.400	 &1.456 &       
4	 &0.411	 &1.225 &       
4	 &0.412	 &1.109 &       
2&  0.417&	 1.010  \\
0.75&	 
8	 &0.408	 &1.713 &       
8	 &0.430	 &1.641 &       
4	 &0.456	 &1.144 &       
4	 &0.466	 &1.072 &       
4&  0.471&	 1.014 \\
0.80&	 
16	 &0.431	 &1.316 &       
8	 &0.479	 &1.116 &       
4	 &0.503	 &1.068 &       
4	 &0.522	 &1.035 &       
4&  0.535&	 1.007 \\
0.85&	 
16	 &0.477	 	&0.850  &	
8	 &0.534	 	&0.864  &	
8	 &0.569	 	&0.871  &	
8	 &0.584	 	&0.919  &	
4&  0.602&	 1.000  \\
0.88&	 
16	 &0.508	 	&0.713  &   
8	 &0.568	 	&0.770  & 
8	 &0.618	 	&0.786  &
8	 &0.641	 	&0.851  &
8&  0.659&	 0.962  \\
0.90&	 
32	 &0.544	 	&0.444  &   
16	 &0.604	 	&0.483  &
8	 &0.652	 	&0.740  &
8	 &0.680	 	&0.812  &
8&  0.704&	 0.948  \\
0.92&	 
32	 &0.584	 	&0.396  &   
16	 &0.645	 	&0.445  &
8	 &0.687	 	&0.701  &
8	 &0.721	 	&0.777  &
8&  0.751&	 0.934  \\
0.94&	 
64	 &0.639	 	&0.226  &   
32	 &0.692	 	&0.252  &
16	 &0.736	 	&0.407  &
8	 &0.763	 	&0.745  &
8&  0.799&	 0.920  \\
0.95&	 
64	 &0.672	 	&0.215  &   
32	 &0.724	 	&0.243  &
16	 &0.763	 	&0.396  &
16	 &0.787	 	&0.429  &
8&  0.824&	 0.913  \\
0.96&	 
128&	 0.712	& 0.123 &   
64	 &0.728	 	&0.172  &   
16	 &0.792	 	&0.386  &
16	 &0.818	 	&0.419  &
8&  0.850&	 0.906  \\
0.97&	 
128&	 0.759	& 0.118 &   
64	 &0.798	 	&0.133  &
32	 &0.827	 	&0.214  &
16	 &0.850	 	&0.409  &
16& 0.876&	 0.545  \\
0.98&	 
256&	 0.818	& 0.066 &   
128&	 0.846	& 0.074 &
32	 &0.867	 	&0.209  &
32	 &0.885	 	&0.219  &
16& 0.912&	 0.534  \\
0.99&	 
1024&	 0.892	& 0.021 &   
256&	 0.908	& 0.040 &
128&	 0.920	& 0.064 &
64	 &0.930	 	&0.117  &
16& 0.949&	 0.525  \\
\end{tabular}
\end{ruledtabular}
\end{table*}
\end{small}
\begin{small}
\begin{table*}[tb]
\caption{Optimal PSPS based on symmetric spatial multiplexing with threshold detectors. Maximal single-photon probabilities $P_{1,\max}$ and the required number of multiplexed units $N_{\opt}$ and input mean photon number $\lambda_{\opt}$ at which they can be achieved, calculated for different $V_r$ router transmissions and five different values of the detector loss $V_D$.\label{tab:binary1}}
\begin{ruledtabular}
\begin{tabular}{c| c@{\;}cc|c@{\;}cc|*2{c@{\;}cc|}c@{\;}cc}
      &\multicolumn{3}{c|}{$V_D=0.3$}&\multicolumn{3}{c|}{$V_D=0.6$}&\multicolumn{3}{c|}{$V_D=0.8$}&\multicolumn{3}{c|}{$V_D=0.9$}&\multicolumn{3}{c}{$V_D=0.98$}\\\hline
$V_r$ &
$N_{\text{opt}}$&$P_{1,\max}$&$\lambda_{\text{opt}}$&
$N_{\text{opt}}$&$P_{1,\max}$&$\lambda_{\text{opt}}$&
$N_{\text{opt}}$&$P_{1,\max}$&$\lambda_{\text{opt}}$&
$N_{\text{opt}}$&$P_{1,\max}$&$\lambda_{\text{opt}}$&
$N_{\text{opt}}$&$P_{1,\max}$&$\lambda_{\text{opt}}$\\\hline
0.30&
4&	 0.369&	 10.900 &
2&	 0.378&	 3.021  &
2&	 0.385&	 2.854  &
2&	 0.385&	 2.814  &
2&	 0.385&	 2.799   \\
0.40&
4	 &0.372	 &5.735&
2	 &0.379	 &2.243&
2	 &0.399	 &2.030&
2	 &0.405	 &1.948&
2&	 0.408&	 1.891\\
0.50&
4	 &0.377	 &3.464&
4	 &0.381	 &3.354&
2	 &0.411	 &1.594&
2	 &0.423	 &1.517&
2&	 0.432&	 1.462\\
0.55&
4	 &0.376	 &2.833&
4	 &0.392	 &2.494&
2	 &0.416	 &1.443&
2	 &0.432	 &1.374&
2&	 0.443&	 1.324\\
0.60&
8	 &0.379	 &3.819&
4	 &0.405	 &1.910&
2	 &0.420	 &1.319&
2	 &0.439	 &1.258&
2&	 0.453&	 1.214\\
0.65&
8	 &0.388	 &2.663&
4	 &0.420	 &1.537&
4	 &0.430	 &1.376& 
2	 &0.446	 &1.162&
2    &0.462&	 1.124\\
0.70&
8	 &0.401	 &1.908&
4	 &0.435	 &1.285&
4	 &0.455	 &1.124&
4	 &0.461	 &1.062&
2&	 0.471&	 1.048\\
0.75&
8	 &0.415	 &1.453&
4	 &0.449	 &1.103&
4	 &0.480	 &0.961&
4	 &0.491	 &0.906&
4&	 0.499&	 0.868\\
0.80&
16	 &0.431	 &1.089&
8	 &0.477	 &0.809&
4	 &0.504	 &0.843&
4	 &0.521	 &0.797&
4&	 0.532&	 0.765\\
0.85&
16	 &0.475	 	&0.750&
8	 &0.520	 	&0.658&
8	 &0.546	 	&0.561&
8	 &0.556	 	&0.525&
4&	 0.565&	 0.689  \\
0.88&
32	 &0.502	 	&0.465  &
16	 &0.551	 	&0.418  &
8	 &0.581	 	&0.510  &
8	 &0.594	 	&0.479  &
8&	 0.604&	 0.458  \\
0.90&
32	 &0.538	 	&0.406  &
16	 &0.585	 	&0.384  &
16	 &0.605	 	&0.322  &
8	 &0.621	 	&0.454  &
8&	 0.632&	 0.435  \\
0.92&
32	 &0.576	 	&0.364  &
16	 &0.620	 	&0.356  &
16	 &0.646	 	&0.300  &
16	 &0.656	 	&0.280  &
16&	 0.663&	 0.267  \\
0.94&
64	 &0.632	 	&0.211  &
32	 &0.672	 	&0.211  &
32	 &0.690	 	&0.175  &
16	 &0.702	 	&0.264  &
16&	 0.711&	 0.252  \\
0.95&
128&	 0.664	& 0.123 &
32	 &0.701	 	&0.204  &
32	 &0.722	 	&0.169  & 
32	 &0.729	 	&0.157  &
16&	 0.735&	 0.246  \\
0.96&
128&	 0.707	& 0.117 &
64	 &0.739	 	&0.119  &
32	 &0.754	 	&0.164  &
32	 &0.763	 	&0.153  &
32&	 0.769&	 0.145  \\
0.97&
256&	 0.754	& 0.066 &
128&	 0.780	& 0.067 &
64	 &0.796	 	&0.095  &
64	 &0.801	 	&0.088  &
64&	 0.805&	 0.083  \\
0.98&
512&	 0.814	& 0.036 &
128&	 0.834	& 0.065 &
128&	 0.844	& 0.053 &
128&	 0.848	& 0.049 &
64&	 0.852&	 0.081  \\
0.99&
1024&	 0.890	& 0.020 &
512&	 0.901	& 0.020 &
256&	 0.907	& 0.029 &
256&	 0.909	& 0.027 &
256&	 0.911&	 0.025\\
\end{tabular}
\end{ruledtabular}
\end{table*}
\end{small}
In Table~\ref{tab:PNRD1} we present a selection of results of the optimization for spatially multiplexed PSPS with SPDs for detector efficiencies $0.3\leq V_D \leq 0.98$ and router transmissions $0.3\leq V_r\leq 0.99$. Here it is assumed that there are no generic losses ($V_b=1$).
For comparison, the corresponding results on the system  equipped with threshold detectors are presented in~Table~\ref{tab:binary1}.
To visualize the results of our optimization, in Fig.~\ref{Figure4} we have plotted the difference $\Delta_P=P_{1,\max}^{\text{SPD}}-P_{1,\max}^{\text{Th}}$ as a function of the detector efficiency $V_D$ and the router transmission $V_r$ over the range $[0.3,1]$ for both parameters. Here $P_{1,\max}^{\text{SPD}}$  is the maximal single-photon probability  that can be achieved with PSPS operated with SPDs, whereas $P_{1,\max}^{\text{Th}}$ is the maximal achievable single-photon probability for PSPS with threshold detectors. 

From the data in Tables~\ref{tab:PNRD1} and~\ref{tab:binary1} and in Fig.~\ref{Figure4} one can conclude that at values $V_r>0.81$ of the router transmission, PSPS based on symmetric spatial multiplexing with SPDs outperform those with threshold detectors, for any detector efficiency under consideration.
In this range a decrease in the detector efficiencies leads to a decrease in the difference in the single-photon probabilities hence at lower detector efficiencies the advantage of using SPDs decreases.
For router transmissions $0.92\leq V_r\leq 0.99$ and detector efficiencies $0.9\leq V_D \leq 0.98$ the difference $\Delta_P$  is at least $0.02$,  while the highest value of the difference $\Delta_P$  is $0.089$ that can be observed at detector efficiency $V_D=0.98$ and router transmission $V_r=0.95$.
We note that at the present state of the art this router transmission is the highest one that seems to be feasible for the ultrafast photon router used in the experiment of Ref.~\cite{Ma2010}.
In the domain of the parameters $V_D$ and $V_r$ bounded by dashed lines in Fig.~\ref{Figure4}, around $V_r\approx 0.8$ , both the PSPS operated with threshold detectors and those with SPDs achieve approximately the same performance, that is, the absolute difference between the single-photon probabilities $|\Delta_P|<0.004$ over the whole range of $V_D$. 
The sign of the difference here depends on the particular values of the parameters $V_D$ and $V_r$.
For router transmissions $0.3\leq V_r<0.78$, SSM based PSPS operated with threshold detectors outperform the ones built with SPDs. Thus there are cases when the use of more advanced detectors does not pay off. 
The physics behind this is that for lower values of the router transmission $V_r$ the incoming signal photons are lost in the multiplexer, therefore the optimal strategy is to allow more than one photon enter the multiplexer in the signal arms at the heralding event. In this range the highest value of the absolute difference $|\Delta_P|$ between the maximal single-photon probabilities is as high as $0.158$, belonging to a detector efficiency of $V_D=0.59$ and router transmission of $V_r=0.3$. 

We note that for detector efficiencies $V_D\to 1$ and router transmissions $0.3\leq V_r<0.6127$ the maximal single-photon probabilities $P_{1,\max}^{\text{SPD}}$ for the systems with SPDs tend to $\exp(-1)$, that is, the maximum single-photon probability of the Poisson distribution.
In this case the optimal number of multiplexed units is $N_{\opt}=1$ (see Table~\ref{tab:PNRD1}) so there is no multiplexing at all. Therefore the maximum single-photon probability is the one that can be achieved by a single heralded nonlinear source.
For detector efficiencies $V_D<1$ the maximal single-photon probabilities $P_{1,\max}^{\text{SPD}}$ are smaller than this value in the considered range of router transmissions. 

\begin{figure}[tb]
\includegraphics[width=\columnwidth]{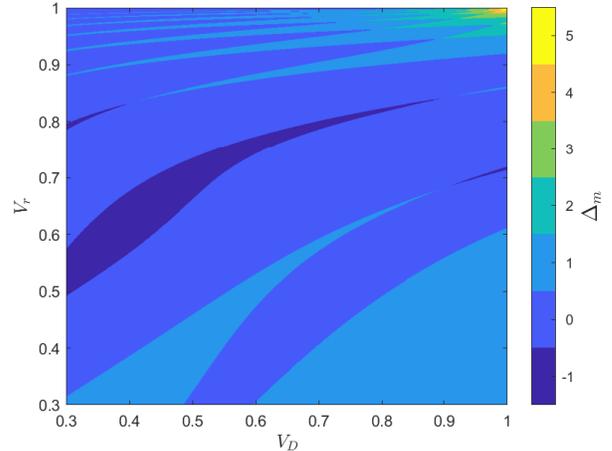}
\caption{\label{Figure5} The difference $\Delta_m=m_{\opt}^{\text{Th}}-m_{\opt}^{\text{SPD}}=\log_2 N_{\opt}^{\text{Th}}-\log_2 N_{\opt}^{\text{SPD}}$ between the optimal number of router levels using SPDs and using threshold detectors, in PSPS based on symmetric spatial multiplexing.}
\end{figure}

From the results in Tables~\ref{tab:PNRD1} and~\ref{tab:binary1} one can also conclude that in the case of both detector types, at low detector efficiencies the optimal number of multiplexed units $N_\opt$ is higher than in the case of high detector efficiency.
In Fig.~\ref{Figure5} we have plotted the difference $\Delta_m=m_{\opt}^{\text{Th}}-m_{\opt}^{\text{SPD}}$ between the optimal number of router levels for systems with SPDs and for the ones with threshold detectors.
The figure shows that in most cases, the optimal number of router levels for SSM based PSPS built with SPDs is less than or equal to the one for the systems with threshold detectors except for a special small range of the loss parameters. 
Hence, the advantage of using the more advanced type of detectors can also appear when it is important to keep the number of units of the system as low as possible. Tables~\ref{tab:PNRD1} and~\ref{tab:binary1} used along with Fig.~\ref{Figure5} can help with implementing this design consideration.

\subsection{Symmetric spatial multiplexing with optimized heralding strategy}\label{sSec:spmods}
In this subsection we consider the optimization of the heralding strategy incorporated into the model described in Sec.~\ref{Sec:math}. 
Let us assume that the PNRDs in the idler arms of the nonlinear sources allow the signal photons to enter the symmetric spatial multiplexer whenever the number of the detected photons is in a prescribed set, which we assume to consist of numbers from $1$ to $J\leq J_b$: $S=\{1,2,\ldots,J\}$. (We shall see later that actually this is the most general form of sets worth considering.)
$J_b$ is the maximal number of photons the PNRDs are capable of distinguishing. Thus the set $S$ is the definition of the heralding strategy, which is uniquely determined by $J$ in the current setting.

\begin{figure}[tb]
\includegraphics[width=\columnwidth]{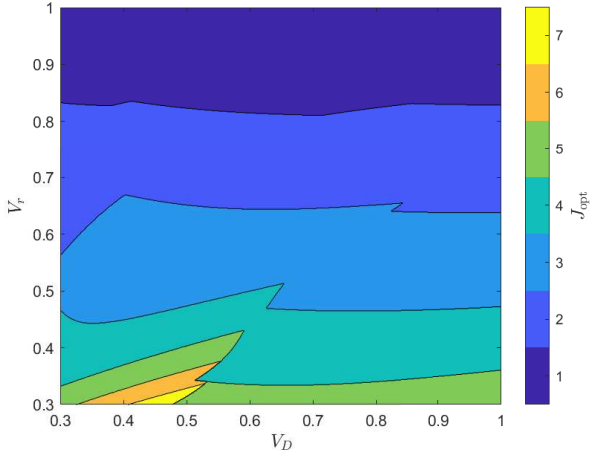}
\caption{\label{Figure6} Optimal maximum accepted photon number $J_\opt$ as a function of the detector efficiency $V_D$ and the router transmission $V_r$ for PSPS based on symmetric spatial multiplexing with PNRDs.}
\end{figure}
We can determine the optimal value of the maximum accepted photon number $J_{\opt}$ by calculating the maximal single-photon probabilities $P_{1,\max}^J$ for all the possible values of the maximum accepted photon numbers $J$ and for detector efficiencies $V_D$ and  router transmissions $V_r$ considered in the previous subsection.
Figure~\ref{Figure6} shows the optimal maximum accepted photon number $J_\opt$ as a function of the detector efficiency $V_D$ and the router transmission $V_r$. 
Decreasing the router transmission $V_r$, that is, increasing the router losses in the multiplexer leads to an increase in the optimal maximum accepted photon number $J_{\opt}$: the optimal heralding strategy gradually shifts from single-photon detection toward threshold detection.
Note, however, that the value of $J_{\opt}$ also depends on the detector efficiency $V_D$.

\begin{figure}[tb]
\includegraphics[width=\columnwidth]{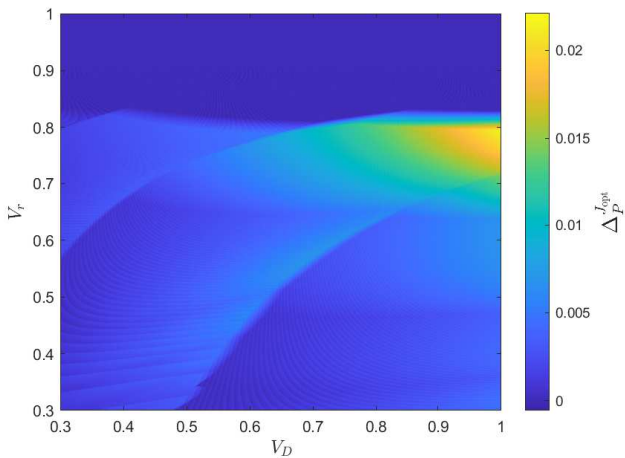}
\caption{\label{Figure7} Difference $\Delta_P^{J_\opt}=P_{1,\max}^{J_\opt}-\sup\left( P_{1,\max}^{\text{Th}},P_{1,\max}^{\text{SPD}}\right)$ as a function of the router transmission $V_r$ and detector efficiency $V_D$, for comparison of the use of an optimal detection strategy in PSPS with symmetric spatial multiplexing, and the scenarios studied in Section~\ref{sec:th}}
\end{figure}

In Fig.~\ref{Figure7} we present the difference $\Delta_P^{J_\opt}=P_{1,\max}^{J_\opt}-\max\left( P_{1,\max}^{\text{Th}},P_{1,\max}^{\text{SPD}}\right)$ as a function  of the detector efficiency $V_D$ and the router transmission $V_r$ over the range $[0.3,1]$ for both parameters.
Here $P_{1,\max}^{\text{SPD}}$ is the maximal single-photon probability when the optimal detection strategy is used, whereas $\max\left( P_{1,\max}^{\text{Th}},P_{1,\max}^{\text{SPD}}\right)$ is the best achievable single-photon probability in either setups (i.e. with threshold or single-photon detectors) studied in Section~\ref{sec:th}.

The figure shows that the parameter domain where using an optimal heralding strategy leads to a relevant enhancement is limited to the router transmissions of $0.65\lesssim V_r\lesssim0.85$ and detector efficiencies of $0.7\lesssim V_D$. 
In this range the optimal heralding strategy consists in letting up to 2 photons enter the system, that is, the optimal maximum accepted photon number $J_{\opt}$ is 2. Nevertheless, this parameter range still extends the domain where the use of PNRDs is a better option than that of threshold detectors. (This domain is the union of the domains in Figs.~\ref{Figure4} and~\ref{Figure7} where the use of PNRDs was found as better.)

Finally, we note that we have considered setups with detection strategies defined by more general sets $S$, e.g. by choosing the set of the accepted photon numbers as e.g. $S=\{2\}$ or $S=\{2,3\}$.  With those choices we found significantly lower single-photon probabilities in the whole range of the parameters, even as compared to single-photon probabilities obtainable by using threshold detectors.

\subsection{Time multiplexers with single-photon and threshold detectors}\label{sSec:btmsptd}

In this subsection we consider PSPS based on time multiplexing schemes described in Sec.~\ref{Sec:overview}. 
Similarly to the case of spatial multiplexing, we compare schemes equipped with SPDs with those containing threshold detectors.
We do this to get a clear picture of the possible benefits of using more advanced detectors.
In this case, however, we restrict our analysis to a selected set of feasible loss parameter values of the optical elements, which are available by state-of-the-art technology.

First we optimize PSPS based on storage loop time multiplexing.
In these systems the optimal operation strategy is to release the latest heralded photon which appeared in the time slot closest to the end of the time period of the periodic source.
Hence, the loss due to the necessary time delay is minimized.
In this case the total transmission probability for the $n$th multiplexed time slot introduced in Eq.~\eqref{eq:LTMLoss} should be modified as follows
\begin{equation}
    V_n^{\text{LTM}}=V_bV_c^n.
\end{equation}
Recently, such a time multiplexed system was realized in Ref.~\cite{Kaneda2019} where a single-photon probability of 0.667 was reported by using 40 time slots.
Motivated by this result, we perform the optimization for the loss parameters of this particular experimental arrangement. 
Accordingly, the transmission corresponding to one cycle in the cavity and the generic transmission coefficient corresponding to switching the heralded photons into the time multiplexer are chosen to be $V_c=0.988$ and $V_b=0.88$, respectively.
We checked that for this set of parameter values single-photon detection is the optimal detection strategy.
\begin{table}[tb]
\caption{Optimal PSPS based on a storage loop time multiplexing with SPDs and threshold detectors. Maximal single-photon probabilities $P_{1,\max}$  and optimal input mean photon number $\lambda_{\opt}$ at which they can be achieved, calculated for different values of the detector loss $V_D$, two values of the number of multiplexed time windows $N$, the transmission of a single cycle of the storage loop $V_c=0.988$, and the general transmission term $V_b=0.88$.}\label{tab:cavity_result_new}
\begin{center}
\begin{ruledtabular}
\begin{tabular}{c|ccc|ccc}
$V_D$&$P_{1,\max}^{\text{Th}}$&$N$&$\lambda_{\opt}^{\text{Th}}$&$P_{1,\max}^{\text{SPD}}$&$N$&$\lambda_{\opt}^{\text{SPD}}$\\\hline
0.60&	0.732& 	 40	&	 0.210&	0.762& 40&	 	 0.267 \\
0.60&	0.737&	 100&	 	 0.186& 0.764&  100& 	 0.248 \\
0.80&	0.759& 	 40	&	 0.186&	0.803& 40&	 	 0.307 \\
0.80&	0.761&	 100&	 	 0.174& 0.803&  100& 	 0.303 \\
0.85&	0.765& 	 40	&	 0.183&	0.814& 40&	 	 0.340 \\
0.90&	0.770& 	 40	&	 0.180&	0.826& 40&	 	 0.398 \\
0.90&	0.771&	 100&	 	 0.171& 0.826&  100& 	 0.397 \\
0.95&	0.775& 	 40	&	 0.177&	0.840& 40&	 	 0.522 \\
0.95&	0.776&	 100&	 	 0.170& 0.840&  100& 	 0.522 \\
0.96&	0.776& 	 40	&	 0.177&	0.844& 40&	 	 0.567 \\
0.97&	0.777& 	 40	&	 0.177&	0.848& 40&	 	 0.625 \\
0.98&	0.778& 	 40	&	 0.176&	0.852& 40&	 	 0.706 \\
0.98&	0.779&	 100&	 	 0.169& 0.852&  100& 	 0.706 \\
\end{tabular}
\end{ruledtabular}
\end{center}
\end{table}
\begin{table}[tb]
\caption{Optimal  PSPS based on a binary bulk time multiplexing with SPDs and threshold detectors. Maximal single-photon probabilities $P_{1,\max}$ and the required number of multiplexed units $N_{\opt}$ and input mean photon numbers $\lambda_{\opt}$ at which they can be achieved, calculated for different detector efficiencies $V_D$, and a fixed set of the loss parameters of the multiplexer, that is, transmission and reflection efficiencies $V_t=0.97$ and  $V_{re}=0.996$, respectively, propagation transmission $V_p=0.95$ and basic generic transmission $V_b=0.996$.}\label{tab:btm_result_new}
\begin{center}
\begin{ruledtabular}
\begin{tabular}{c|ccc|ccc}
$V_D$&$P_{1,\max}^{\text{Th}}$&$N_{\opt}^{\text{Th}}$&$\lambda_{\opt}^{\text{Th}}$&$P_{1,\max}^{\text{SPD}}$&$N_{\opt}^{\text{SPD}}$&$\lambda_{\opt}^{\text{SPD}}$\\\hline
0.60&0.838	& 256	&  0.042 &0.849	& 128	  &0.080 \\
0.80&0.847	& 128	&  0.058 &0.868	& 64	  &0.130 \\
0.85&0.849	& 128	&  0.056 &0.874	& 32	  &0.226 \\
0.90&0.851	& 128	&  0.054 &0.883	& 32	  &0.241 \\
0.95&0.853	& 128	&  0.053 &0.895	& 16	  &0.481 \\
0.96&0.853	& 128	&  0.053 &0.898	& 16	  &0.506 \\
0.97&0.854	& 128	&  0.052 &0.902	& 16	  &0.542 \\
0.98&0.854	& 128	&  0.052 &0.907	& 16	  &0.600 \\
\end{tabular}
\end{ruledtabular}
\end{center}
\end{table}

Table~\ref{tab:cavity_result_new} presents the results of the optimization for SPDs and threshold detectors for various values of the detector efficiency $V_D$.
For the described operation strategy (i.e. releasing the last heralded photon) the achievable single-photon probability $P_{1}$ increases and saturates with increasing the number of time slots $N$, therefore one cannot find a particular optimal value $N_{\opt}$ for this quantity.
Hence, we assumed $N=40$ time slots, as it has been realized in the cited experiments.
We also show results for $N=100$ in order to demonstrate the saturation of the single-photon probability.
From the results it is clear that single-photon sources equipped with SPDs outperform sources equipped with threshold detectors for the considered set of loss parameters.
The enhancement in the single-photon probability is bigger at high detector efficiencies. 

We note that the data of Table~\ref{tab:cavity_result_new} were obtained assuming a Poissonian distribution for the number of generated photon pairs in order to compare the results of the optimization with the those of other multiplexers presented previously.
However, in case of strong spectral filtering at the heralding stage the number of generated photon pairs follows a thermal distribution, as it is the case described in Ref.~\cite{Kaneda2019}.
Hence, we carried out the optimization of the same system, with the same loss parameters, for $N=40$ and  $V_D=0.6$, and under the assumption of thermal instead of Poissonian input distribution this time. This  resulted in a single-photon probability $P_{1,\max}^{\text{SPD}}=0.713$, which corresponds to the ones achieved in the reported experiments. In the cited experiment, however, the photon number resolving detection was realized with spatially multiplexed detectors. Considering the same arrangement with an inherent PNRD of efficiency as high as $V_D=0.98$~\cite{Fukuda2011} and with $N=100$ time slots offers a single-photon probability $P_{1,\max}^{\text{SPD}}=0.829$.

Finally, let us turn our attention to the analysis of PSPS based on binary bulk time multiplexing described in Sec.~\ref{Sec:overview}.
Table~\ref{tab:btm_result_new} contains results of the optimization of such systems for state-of-the-art loss parameters, that are the transmission efficiency of $V_t=0.97$, the reflection efficiency of $V_{\rm re}=0.996$, and the propagation transmission of $V_p=0.95$ \cite{Adam2014}. Also in this case, we compare the use of PNRDs with that of threshold detectors.

Similarly to the previously analyzed system, we found that single-photon detection is the optimal detection strategy for this particular set of parameters.
These results clearly show that such PSPS built with single-photon detectors have higher achievable single-photon probabilities than the same constructions built with threshold detectors.
As the detector efficiency increases, the difference between the single-photon probabilities achieved with threshold detectors and those achieved with single-photon detectors increases as well.
In these systems the optimal number of multiplexed time slots $N_{\opt}$ is significantly lower when PNRDs are applied.
From Tables~\ref{tab:PNRD1}-\ref{tab:btm_result_new} one can conclude that from amongst all multiplexed PSPS considered in this paper, in the case of experimentally feasible values of the loss parameters, the system with binary bulk time multiplexing and single-photon detectors offers the highest single-photon probability which is $P_{1,\max}^{\text{SPD}}=0.907$.

\section{Conclusions}\label{Sec:conc}

We gave a full statistical description of spatial or time multiplexing based periodic single-photon sources realized with photon-number-resolving detectors. This model facilitates the optimization of the system by determining the optimal system size and input photon numbers in order to achieve a maximal single-photon probability at the output, given the losses of the system and the photon number distribution of the source to be heralded. 

Our analysis of the symmetric spatial multiplexing-based schemes of this kind we have found that the use of the more advanced photon-number-resolving detectors do not necessarily lead to an improvement as compared to the use of threshold detectors; we have identified the range where an improvement has been achieved. We also found that this range can be further extended by introducing an appropriate heralding strategy defined in terms of detected photon numbers at the detectors.

The analysis of time multiplexed arrangement has been focused on particular value sets of the parameters which are either typical in current experimental realizations or would at least be feasible at the current state of art. In case of these systems we have found that photon number resolving detectors always offer an improvement. The optimal approach is actually the detection of exactly one photon. Modeling a particular recent experiment based on time-loop multiplexing, the results of our detailed analysis were inline with the experimentally observed ones, and we have assessed the potential further improvement via optimization. In an other case, that for binary bulk time multiplexing, we have found that if the application of an efficient inherently photon-number-resolving detector were feasible, the single-photon probability could reach a value as high as $0.907$, the best that we have found amongst all the studied systems.

Our analysis leads to a detailed understanding of an important line of experimental arrangements: multiplexed periodic single photon sources, which are essential ingredients of many applications. Our quantitative analysis can serve as a guidance for the optimal design of these.

\acknowledgments{This research was supported by the National Research, Development and Innovation Office, Hungary (Projects No. K124351 and No. 2017-1.2.1-NKP-2017-00001 HunQuTech) and by the European Union (Grants No. EFOP-3.6.1.-16-2016-00004, No. EFOP-3.6.2-16-2017-00005, and No. EFOP-3.4.3-16-2016-00005).}
%

\end{document}